**Defect Landscape Engineering Suppresses Helium Damage in Ceramics**

Nabil Daghbouj[1]*, Ahmed Tamer AlMotasem[1]*, Bingsheng Li[2]*, Vladimir Krsjak[3], Jan Duchoň[4], Fang. Ge[5], Maceig Oskar Liedke[6], Andreas Wagner[6], Mohamed Bensalem[1], Fateh Bahadur[1], Frans Munnik[7], Miroslav Karlik[8], Anna Macková[9,10], Tomas Polcar[1,11], William J. Weber[12]

[1]*Department of Control Engineering, Faculty of Electrical Engineering, Czech Technical University in Prague, Technická 2, 160 00 Prague 6, Czechia*

[2]*State Key Laboratory for Environment-friendly Energy Materials, Southwest University of Science and Technology, Mianyang, Sichuan 621010, China*

[3]*Institute of Nuclear and Physical Engineering, Faculty of Electrical Engineering and Information Technology, Slovak University of Technology, Ilkovicova 3, 812 19 Bratislava, Slovakia*

[4]*Institute of Physics of the Czech Academy of Sciences, Na Slovance 1999/2, 182 21 Prague 8, Czechia*

[5]*Laboratory of Advanced Nano Materials and Devices, Ningbo Institute of Materials Technology and Engineering, Chinese Academy of Sciences, Ningbo 315201, China*

[6]*Institute of Radiation Physics, Helmholtz-Zentrum Dresden-Rossendorf, Bautzner Landstr. 400, 01328 Dresden, Germany*

[7]*Helmholtz-Zentrum Institute of Ion Beam Physics and Materials Research, Helmholtz-Zentrum Bautzner Landstr. 400, 01328 Dresden, Germany*

[8]*Department of Materials, Faculty of Nuclear Sciences and Physical Engineering, Czech Technical University in Prague, Trojanova 13, 120 00 Prague 2, Czechia*

[9]*Nuclear Physics Institute of the Czech Academy of Sciences, 250 68 Husinec-Řež, Czechia*

[10]*Department of Physics, Faculty of Science, University of J. E. Purkyně, 400 96 Ústí nad Labem, Czech Republic*

[11]*School of Engineering, University of Southampton, Southampton SO17 1BJ, United Kingdom*

[12]*Department of Materials Science & Engineering, University of Tennessee, Knoxville, TN 37996, USA*

**Abstract**

Helium accumulation in structural ceramics used in nuclear, fusion, and aerospace systems causes swelling, cracking, and early failure, yet controlling this damage has remained elusive. Here, we introduce defect landscape engineering, the deliberate creation of vacancy clusters prior to helium exposure, as a general strategy to suppress helium-induced degradation. Using α-SiC as a model, we combine advanced microscopy, strain mapping, helium depth profiling, positron annihilation spectroscopy, and atomistic simulations to demonstrate that tailored pre-damage transforms helium defect evolution. Instead of forming extended platelets and nanocracks, helium is trapped in stable, uniformly dispersed nanobubbles. Simulations reveal that small vacancy clusters act as dual-function sinks for irradiation-induced interstitials and preferential helium traps, fundamentally altering cascade recombination dynamics. This mechanism is composition-independent and scalable, offering a new design principle for radiation-tolerant ceramics across carbides, nitrides, and oxides. By viewing defect control as a tunable parameter instead of a fixed material property, this work outlines a possible design route toward enhanced radiation tolerance in ceramics used in extreme environments.

**Introduction**

As the world races to decarbonize energy systems and mitigate the intensifying effects of climate change, advanced nuclear energy technologies have emerged as a critical solution. Their potential to deliver reliable, low-carbon power is unparalleled, but realizing this promise hinges on one key factor: the resilience of structural materials operating in some of the most extreme environments engineered by humans. Inside nuclear reactor cores, materials face relentless bombardment by high-energy neutrons, reaching doses as high as 200 displacements per atom (dpa) [1,2]. Under such intense conditions, maintaining the mechanical integrity, safety, and long-term performance of these materials becomes one of the most pressing challenges in the field.

The extension of current reactor lifespans and the deployment of next-generation fission and fusion systems further intensify the demand for materials with exceptional irradiation tolerance [3,4]. When exposed to neutron or heavy-ion irradiation, metals and ceramics accumulate a spectrum of defects, vacancies, interstitials, dislocation loops, voids, bubbles, and stacking fault tetrahedra [5–11], which evolve into macroscopic degradation phenomena such as embrittlement, void swelling, radiation-induced hardening, and creep [12–17].

A particularly insidious contributor to irradiation damage is helium, produced by transmutation reactions or introduced externally in plasma-facing environments. Because of its extremely low solubility in solids, helium atoms readily cluster to form bubbles, cavities, and platelets, which can grow and coalesce into nanocracks and larger-scale defects [18–22]. As these gas-filled cavities accumulate beneath material surfaces, the internal pressure can reach a critical threshold, leading to blistering, surface upheaval, and ultimately, spallation [23–25]. This blistering behavior has been observed in materials like silicon carbide (SiC), where helium implantation leads to bubble nucleation, platelet growth, and surface failure [26–28].

Developing materials with built-in resistance to such degradation is now a central objective of nuclear materials science. A wave of innovation has yielded several promising strategies such as grain boundary engineering, which enhances defect sink efficiency by tailoring the grain boundary character and distribution [29–32]; Multilayered nanocomposites, which leverage interface-dominated architectures to absorb and annihilate radiation-induced defects [33–36]; Metallic glasses, whose amorphous structure resists amorphization and void formation under irradiation [37–39]; Nanochannel structures, created via physical vapor deposition (PVD), which promote helium outgassing and thermal self-healing [40,41]; Nanotwinned metals, where coherent twin boundaries serve as defect traps and enhance both strength and radiation resistance [42,43];

High-entropy alloys (HEAs), which benefit from sluggish diffusion and lattice distortion effects, enabling greater radiation tolerance than conventional alloys [44–46].

Among ceramic materials, SiC remains one of the most promising candidates for structural components in both fission and fusion environments. Its high thermal conductivity, corrosion resistance, and radiation stability make it an ideal material for cladding and structural applications. However, SiC is not immune to helium-induced damage. Due to its low helium solubility and diffusion, implanted He atoms become trapped in vacancies or pre-existing defects, forming stable clusters that evolve into platelets and eventually blisters [47]. The immobility of helium in such a matrix makes it exceptionally difficult to mitigate damage once initiated.

Understanding helium transport and retention in SiC, especially in the presence of irradiation-induced damage, is therefore of urgent scientific and technological interest. However, the fundamental properties of helium behavior in solids, such as diffusion and resolution energies, remain poorly characterized under realistic reactor conditions. A recent study by Agarwal et al. [48] introduced a novel approach using pre-damaged Fe subjected to self-ion irradiation. By combining Doppler Broadening Spectroscopy-Positron Annihilation Lifetime Spectroscopy (DBS-PALS) with transmission electron microscopy, they demonstrated that the presence of initial voids could influence cascade damage, reduce the size of large voids while increasing small vacancy clusters, highlighting a previously unrecognized mechanism of defect evolution. Inspired by this insight, here, we introduce defect landscape engineering, the deliberate tailoring of vacancy cluster populations and distributions, as a strategy to control He behavior at the atomic scale. By designing pre-existing damage profiles, it is possible to redirect helium transport, stabilize nanoscale bubbles, and suppress extended defect formation. Using α-SiC as a model ceramic, we demonstrate that pre-existing vacancy clusters, introduced by carbon self-ion implantation to defined damage levels (0.25 dpa and 0.36 dpa) to simulate reactor-relevant defect populations, act as both sinks for cascade-generated interstitials and preferential traps for He. This dual functionality alters the thermodynamics and kinetics of He clustering, resulting in dispersed, stable nanobubbles instead of large platelets and cracks.

Our integrated experimental, computational approach combines bright-field STEM, nano-precession electron diffraction strain mapping, elastic recoil detection analysis, depth-resolved positron annihilation spectroscopy, and atomistic simulations to uncover the mechanisms linking engineered defect landscapes to He damage suppression. Our findings establish a composition-independent and scalable route for tuning radiation response in ceramics. The demonstrated control over He defect evolution provides a design principle applicable to a wide class of radiation-sensitive materials, enabling enhanced performance in nuclear, fusion, and other extreme-environment applications.

**Results**

Helium Behavior as a Function of Damage and Temperature

To evaluate the role of engineered vacancy landscapes in controlling helium defect evolution, we implanted α-SiC with carbon ions to introduce controlled levels of pre-existing damage prior to helium irradiation. The virgin α-SiC consists of larger grains, as demonstrated by the STEM-EDS in Supplementary Figure 1. The carbon ion energies and fluences were chosen to produce damage peaks at depths overlapping or slightly deeper than the helium implantation range, ensuring strong spatial interaction between helium atoms and defect-rich zones. Fig. 1 presents Elastic Recoil Detection Analysis (ERDA)- derived helium depth profiles under three conditions: (i) He implantation at room temperature (RT, black curve), (ii) He implantation at 750 °C (red curve), and (iii) He implantation at 750 °C into SiC pre-damaged by carbon ions to 0.36 dpa (blue curve). These are compared with Stopping and Range of Ions in Matter (SRIM)- simulated depth profiles for He and C ions. In undamaged samples implanted at RT and 750 °C, two distinct helium peaks appear at approximately 850 nm and 960 nm, suggesting partial redistribution and thermal diffusion. The experimental profiles deviate from SRIM predictions, indicating temperature-dependent He mobility. In the pre-damaged sample, the He profile is slightly shallower, with a dominant peak near 944 nm, closely matching the SRIM-predicted He-ion profile. By contrast, when implanted alone at RT or 750 °C, the dominant peak occurs deeper, at ~960 nm, than the SRIM prediction. These results highlight the combined influence of temperature and SiC matrix condition on He transport. The observed partial redistribution of He toward shallower depths may be linked to excess vacancy concentrations just below the damage peak. Such vacancies likely arise from forward-scattered recoils that become interstitials, leaving behind vacancy-rich zones capable of trapping helium. The apparent higher He concentration near the end of the range is likely an artefact caused by C recoil ions entering the detector, due to ion beam energies exceeding the optimal range for the entrance foil.

**Helium-Induced Microstructural Evolution at Elevated Temperature**

Helium implantation at 750 °C leads to pronounced microstructural changes in sintered α-SiC, as observed through conventional TEM analysis of transverse FIB-prepared foils (Fig. 2). The implanted samples exhibit well-defined helium platelets exceeding 20 nm in length (marked with red arrows), along with nanocracks up to 120 nm, highlighted by yellow arrows in Fig. 2f.

These features spatially coincide with the peak helium concentration, as confirmed by overlaid ERDA and SRIM depth profiles. Lattice-resolution TEM imaging (Fig. 2g) reveals that the platelets are nearly filled with helium gas trapped between distorted atomic planes, providing direct evidence of gas accumulation and localized structural degradation. Importantly, helium implantation at this elevated temperature does not result in amorphization, even at a peak damage level of 1.26 dpa. In contrast, implantation at RT induces clear amorphization, and large He bubbles are formed as shown in Supplementary Figure 2. BF-TEM images recorded under both under- and over-focused conditions (Figs. 2b and c) reveal a grain boundary (GB) within the implantation zone that remains free of helium bubbles and platelets. Instead, a bubble-denuded zone (BDZ) is observed along the GB, indicating that this interface acts as an efficient sink for irradiation-induced point defects and/or small clusters, thereby suppressing bubble nucleation in its vicinity. Similar BDZ formation mechanisms, involving defect annihilation or preferential trapping at extended defects such as grain boundaries, have been reported in a range of irradiated materials [49,50]. However, the surrounding matrix is visibly distorted, indicating the presence of helium-induced defects.

To quantify the resulting lattice distortion, nano-beam precession electron diffraction (N-PED) was performed. The corresponding strain map (Fig. 2d) was generated by aligning the sample to the [214] zone axis and calculating atomic plane deviations relative to a reference region of pristine, unstrained SiC (see Supplementary Figure 3). The map reveals localized strain, with values reaching approximately 2.5% near and along the GBs. Within the grain interior, a peak tensile strain of nearly 1% aligns with the region of highest helium concentration. The strain profile shown in Fig. 2e was extracted from the region outlined in the strain map. Further analysis of strain evolution and defect behavior near the GBs is discussed elsewhere [51].

**Effect of Low-Dose Pre-Damage on Helium-Induced Microstructure**

Fig. 3a presents a BF-STEM image of sintered α-SiC irradiated with carbon ions at RT to induce a low pre-damage level of 0.25 dpa, with the damage peaking at approximately 1060 nm. The superimposed damage profile calculated from SRIM highlights this peak, and a high-resolution inset shows the region with the highest defect density. STEM-EELS (Supplementary Figure 4) shows an almost similar C and Si ratio even at the peak C concentration, due to the low carbon content (0.2 at%) and the resolution limit of the spectrometer. Radiation-induced defects from carbon implantation lead to localized damage accumulation, particularly in the area of peak displacement. Ion implantation typically causes lattice distortion, primarily tensile and oriented perpendicular to the grain surface [22,26]. Using the same [214] zone axis as in the He-only sample, strain mapping via N-PED (Fig. 3b) reveals that the maximum tensile strain reaches nearly 3% at the peak of the carbon-induced damage. The corresponding strain profile is shown in Fig. 3c, and more details about strain measurements are shown in Supplementary Figure 5. HRTEM along the [$\bar{2}$203]zone axis (Fig. 3d) shows a heavily defective microstructure, and the corresponding FFT (inset) highlights the (11$\bar{2}$0)Reflection is used for geometric phase analysis. The inverse-FFT filtered image (Fig. 3c) reveals well-defined (11$\bar{2}$0)fringes that are locally bent or discontinuous, indicating lattice distortion. This is captured more clearly in the phase map for g = (11$\bar{2}$0)(Fig. 3d), where the phase varies almost linearly in relatively undisturbed regions but exhibits pronounced kinks and necklace-like perturbations along several curved bands. The strain tensor components $\varepsilon_{xx}$ and $\varepsilon_{yy}$ derived from this phase (Supplementary Figure 6) show corresponding networks of alternating tensile and compressive regions, reflecting strong local lattice distortion associated with a high density of irradiation-induced defect clusters and local lattice bending. The same sample was subsequently implanted with helium at 750 °C at a fluence of $1 \times 10^{17}$ He/cm². The resulting microstructure is shown in the BF-TEM image in Fig. 4a, and STEM-EDS (Supplementary Figure 7), with both the helium and carbon damage profiles superimposed. A magnified view of the irradiated zone (highlighted by a yellow circle in Fig. 4a) is provided in Fig. 4b. This region reveals the formation of discrete helium bubbles (yellow arrows) and aligned bubble arrays (blue arrows), with no evidence of helium platelet formation. These bubble arrays are more prominent near the edges of the implantation zone, where overall bubble density is lower. The presence of helium bubbles at the tail end of the damaged region may be influenced by local strain gradients and helium diffusion toward the most strained areas. This suggests a helium redistribution mechanism driven by defect density and strain fields, which effectively suppresses the formation of extended platelets and nanocracks commonly observed in He-only implanted samples. To perform HRTEM analysis, a wide-area image of the grain oriented along the zone axis is presented in Fig. 4c and 4d. High-resolution TEM images were taken near the surface (Fig. 4c) and at the peak damage region (Fig. 4d), which are highlighted by red and green circles, respectively. The microstructure at the peak damage (Fig. 4d) exhibits a heavily defective and distorted lattice, indicating significant radiation-induced damage compared to the near-surface region.

**Effect of High-Dose Pre-Damage on Helium platelets, bubble array Suppression**

Fig. 5a shows a BF-STEM image of sintered α-SiC irradiated with carbon ions to achieve a higher pre-damage level of 0.36 dpa, with the damage peak located around 1060 nm. The superimposed profiles reveal a broader and more concentrated damage region, extending from approximately 850 nm to 1200 nm, compared to the 0.25 dpa condition. STEM-EDS analysis is presented in Supplementary Figure 8. HRTEM imaging along the $[\bar{2}203]$ zone axis (Fig. 5b) reveals a strongly defected microstructure, and the corresponding FFT (inset) shows the $(11\bar{2}0)$ reflection selected for geometric phase analysis. The inverse-FFT filtered image (Fig. 5c) displays well-defined $(11\bar{2}0)$ fringes that become locally bent or discontinuous, evidencing lattice distortion. This behavior is emphasized in the phase map for $g = (11\bar{2}0)$ (Fig. 5d), where the phase is nearly linear in relatively undisturbed areas but shows pronounced kinks and necklace-like perturbations along several curved bands. The strain tensor components $\varepsilon_{xx}$ and $\varepsilon_{yy}$ derived from this phase (Supplementary Figure 9) exhibit corresponding networks of alternating tensile and compressive regions, indicative of strong local lattice distortion caused by a high density of irradiation-induced defect clusters and local lattice bending.

After pre-damaging, the sample was co-implanted with helium at 750 °C using the same fluence of $1 \times 10^{17}$ He/cm$^2$ as performed previously. The resulting microstructure is shown in Fig. 6a, with both helium and carbon ion profiles overlaid, and the STEM-EDS is shown in Supplementary Figure 10.

A magnified view of the region marked by the red circle (He peak concentration) and the blue circle (C peak concentration, which appears with black contrast) in Fig. 6a is shown in Figs. 6c and d, captured under-focused conditions, respectively. These images reveal only nanoscale helium bubbles, approximately 1 nm in diameter (indicated by yellow arrows), with no evidence of helium platelets. Strain mapping for the co-irradiated sample is presented in Fig. 6c, using the same [214] zone axis as in previous measurements to ensure consistency. The maximum tensile strain is approximately 1.5%, located at the trailing edge of the carbon concentration peak, visible as a ~60 nm thick black layer (black contrast in a), corresponding to the intense red region in the strain map. This strain level is higher than that induced by helium-only implantation but significantly lower than the 3% observed in the 0.25 dpa pre-damaged sample. The pronounced strain at the tail of the carbon profile suggests that, during the second irradiation step with helium ions at 750 °C, some carbon atoms diffused slightly deeper into the lattice and possibly formed interstitial clusters, contributing to localized strain. These observations indicate that higher pre-damage levels not only influence the redistribution of implanted helium but also mitigate strain accumulation by promoting the formation of isolated nanoscale bubbles instead of extended defects such as helium platelets or nanocracks. A grain aligned with the [214] zone axis and its corresponding SADP are shown in Fig. 6e. High-resolution TEM images were acquired from a region close to the surface (Fig. 6f) and from the projected damage peak (Fig. 6g), marked by red and green circles, respectively. The microstructure at the damage peak (Fig. 4d) displays a strongly distorted, defect-rich lattice, evidencing much more severe irradiation damage than in the near-surface region.

**Comparative Strain Behavior Across Irradiation Conditions**

Fig. 7 presents a comparative analysis of strain profiles along the [01-1-4] direction measured perpendicular to the grain surface under three different irradiation conditions: helium implantation at 750 °C without pre-damage (red curve), carbon pre-damage to 0.25 dpa followed by helium implantation at 750 °C (green curve), and carbon pre-damage to 0.36 dpa followed by helium implantation at 750 °C (blue curve). In the He-only sample (red), the induced strain in the grain interior is moderate, peaking at approximately 1%. With the introduction of 0.25 dpa carbon-induced pre-damage (green), the strain significantly increases to nearly 3%, due to enhanced lattice distortion near the defect-rich region. However, when the pre-damage level is raised to 0.36 dpa (blue), the peak strain is reduced to ~1.7%. In the 0.36 dpa condition, the maximum strain aligns with the tail of the carbon concentration rather than the helium concentration peak. This suggests that carbon interstitials contribute more strongly to lattice distortion and that helium may have partially diffused toward the pre-damaged region. The reduction in strain compared to carbon-only damage indicates that some defects may be annihilated during helium implantation at 750 °C, as the diffusion of C and Si interstitials in SiC becomes active above 600 °C [52]. Despite the lower peak strain, the strain distribution becomes broader, implying that helium atoms are being trapped in pre-existing or irradiation-induced vacancies, forming $He_nV_m$ clusters. This clustering leads to widespread lattice distortion while reducing the availability of mobile helium and vacancies that could otherwise contribute to larger-scale defect growth. Overall, these results reveal a non-linear relationship between pre-damage level and strain evolution, suggesting the existence of an optimal defect density that minimizes helium-induced swelling and mechanical degradation.

**Positron Annihilation Spectroscopy Analysis of Defect Structures**

Doppler Broadening Spectroscopy (DBS) using a slow positron beam, complemented by VEPFIT modeling, is used to investigate defect structures in sintered SiC under varying irradiation conditions. This technique leverages the fact that positrons are preferentially trapped at lattice defects, particularly vacancies, due to the absence of positively charged ion cores. Once trapped, the fraction of positrons that annihilate with valence electrons results in a narrower energy distribution of the 511 keV gamma-ray peak. The width of this distribution is quantified by the S parameter, a lineshape parameter representing annihilation with low-momentum electrons and serving as a sensitive indicator of defect density. To assess the initial depth distribution of implanted positrons, stopping profiles were calculated using the Makhovian distribution function [53] and are shown in Fig. 8 for incident positron energies of 8, 16, 24, and 35 keV. Stopping profiles of monoenergetic positrons and implanted ions in α-SiC, illustrating the depth-dependent sensitivity of slow positron beam analysis. The profiles include 300 keV He ions, 1080 keV C ions, and positrons at varying implantation energies. For He and C ions, the total stopping power (i.e., nuclear + electronic energy loss) is considered. For positrons, the profiles reflect the electronic (inelastic) stopping power only, as is typical for positron implantation simulations. After implantation, positrons quickly lose energy through inelastic scattering, diffuse within the lattice, and eventually annihilate. The depth of positron penetration, as determined by the Makhov function, correlates directly with the incident positron energy [54]. As observed in the stopping profiles, lower implantation energies result in shallow penetration depths, while higher energies allow for deeper probing, albeit with reduced depth resolution. The 35 keV positron energy not only overlaps with both the 300 keV He and 1080 keV C ion implantation zones, but significantly exceeds the ion beam modified region, effectively probing into the unirradiated bulk. However, despite this deeper probing, the measured S parameter does not return to the background bulk level, as the positrons still 'sense' the defect-rich implanted zone during their thermal diffusion and annihilation, leading to an extended influence of the irradiation-modified layer on the positron response

Fig. 9a presents the S parameter profiles, S(E), plotted as a function of positron energy and corresponding mean implantation depth. The S parameter reflects the proportion of positrons annihilating with low-momentum valence electrons and is proportional to the overall defect concentration. The results reveal marked differences in S(E) behavior between samples irradiated solely with He and those pre-implanted with C ions followed by He irradiation, indicating distinct defect structures. A key advantage of variable-energy DBS is its ability to extract information about positron diffusion within the material, which is used to estimate defect-specific trapping characteristics. Positron diffusion lengths ($L^+$) were extracted by fitting the S(E) profiles using the VEPFIT code, and the results are shown in Fig. 9b. VEPFIT, a standard tool in positron annihilation studies [55], models depth-dependent S parameter profiles by solving the time-averaged positron density equation across multiple layers.

Table 1 summarizes the results of the fitting procedure based on a four-layer model. The table presents the obtained S parameters and positron diffusion lengths for each layer, along with parameters that were fixed during the fitting process to ensure both the numerical stability of the iteration and the physical plausibility of the results. The fitted layer structure and corresponding positron diffusion lengths are schematically illustrated in Fig. 10.

The fitted layer model, illustrated in Fig. 10, offers the following interpretation of the depth-resolved positron annihilation characteristics in He-irradiated and C+He co-irradiated α-SiC. Each layer corresponds to a physically distinct region within the sample, with characteristic positron diffusion lengths and S parameters that reflect the underlying defect structure and its evolution under different irradiation conditions.

i) Layer 1 (~10 nm): Surface-related features due to oxidation or sample preparation. Though included for fitting accuracy, it has little physical relevance.
ii) Layer 2 (550–900 nm): Ion track region where displacement damage increases the S parameter, indicating enhanced positron trapping. The extent of damage and trapping efficiency depends on irradiation temperature and prior implantation.
iii) Layer 3 (implantation/damage peak): Displays the most prominent differences.
    o In the RT He-irradiated sample: High S and short diffusion length (~10 nm) suggest dense, small defect clusters.
    o In the 750 °C He-only sample: Lower S and longer diffusion length (~70 nm) indicate fewer but larger defects, such as coalesced platelets or bubbles.
    o In the C + He co-implanted sample: Intermediate S and diffusion length (~50 nm), consistent with small, dispersed vacancy-type defects, as also observed in TEM images showing nanoscale bubbles but no large platelets.
iv) Layer 4 (bulk SiC): Represents undamaged SiC with a fixed S parameter and diffusion length of 85 nm, based on literature values.

Overall, the DBS results are in strong agreement with TEM observations. High-temperature He implantation alone tends to produce fewer but larger defects, such as extended platelets or bubbles. In contrast, pre-damaging the material via C implantation alters the defect evolution pathway, promoting the formation of smaller, more numerous vacancy-type defects that effectively trap helium and inhibit the development of larger defect structures. The co-implanted sample exhibits a higher degree of positron trapping than the He-only sample, indicating a greater population of small vacancy clusters. This interpretation is further corroborated by strain analysis and simulation results, reinforcing the role of pre-damage in tailoring the defect landscape [55].

**Discussion**

The results demonstrate that pre-damage engineering in Sintered α-SiC significantly alters helium-induced defect evolution. Instead of forming extended defects such as platelets and nanocracks, the microstructure transitions toward stable, sub-nanometer helium-vacancy clusters. This transformation is driven by the presence of pre-existing voids and vacancy clusters, introduced via controlled carbon self-ion irradiation, which act as efficient sinks for both interstitials and helium atoms. As shown in Fig. 1, helium implanted at 750 °C into undamaged α-SiC undergoes limited redistribution and readily nucleates extended defects, most notably, nanocracks and platelets localized near the peak helium concentration (~930 nm), as observed in Figs. 3 a-g. In contrast, samples that underwent co-irradiation with carbon ions at doses of 0.25-0.5 dpa exhibit a marked shift in helium migration toward deeper regions (~1120 nm), aligning with the carbon-induced damage peak. This behavior confirms preferential helium trapping in defect-rich zones.

The effect of increasing pre-damage levels is evident in Figs. 4 and 5. At 0.25 dpa, helium forms discrete bubbles and aligned bubble arrays, reflecting partial suppression of extended defects. At 0.5 dpa, no platelets/nanocracks are observed; only isolated, sub-nanometer helium bubbles remain, indicating complete suppression of large-scale defect formation.

To elucidate the atomic-level mechanisms behind these observations, we conducted a molecular dynamics (MD) cascade simulation of 6H-SiC with a preexisting void. The primary knock-on atom (PKA), typically a Si atom, is given 10 keV kinetic energy directed along the negative z-axis. In all cascade simulations, the initial distance between the PKA atom to the center of the void is about 15 Å for the sake of comparison. Fig. 11 presents atomistic snapshots for both pristine (Fig. 11a) and pre-damaged systems containing voids of varying radii (5, 10, 15, and 20 Å; Figs. 11 b-e). At the final stage of cascade simulations, we performed a cluster analysis to investigate the impact of pre-existing voids on the cascade defects. The snapshots, Figs. 11 a-e, the largest vacancy-clusters are highlighted in blue, while red circles denote the pre-existing voids. In general, we noticed that the number of $V_C$ is significantly higher than that of $V_{Si}$ this behavior has been recently reported by [56], ensuring the suitability and correctness of the selected interatomic potential. This behavior can be attributed to the lower formation energy of $V_C$ (4.52 eV) compared to (8.03) for $V_{Si}$ in 6H-SiC as predicted by DFT calculations [57]. Furthermore, we compared the size of the cascade defects, namely vacancy clusters excluding the pre-existing void, and the result is presented in Fig. 11f. The data reveal a clear inverse relationship between pre-existing void and cascade defects, i.e., as the initial void size increases, the formation of large vacancy clusters is increasingly suppressed. This trend aligns with earlier studies that reported reduced defect generation in pre-damaged materials [58,59]. Supporting this, Ayanoglu et al. [60] experimentally observed void shrinkage in austenitic Fe-21Cr-32Ni alloy foils pre-irradiated with 5 MeV $Fe^{++}$ ions and subsequently re-irradiated in situ with Kr ions over a temperature range of 50-713 K.

In these studies, the suppression of newly formed large vacancy clusters was attributed to the presence of pre-existing damage, though the underlying mechanism remained unclear. In this work, we propose a mechanism for void shrinkage based on the interaction of newly generated interstitials and Frenkel pairs with pre-existing voids (see Supplementary Movie 1). Fig. 12 illustrates the microstructural evolution of defects during both the intermediate and final stages of the cascade process, highlighting the dynamic interplay between cascade-induced defects and pre-existing void structures. Various types of defects are generated, including Carbon vacancies (Vc), silicon vacancies ($V_{Si}$), silicon interstitials ($I_{Si}$), and carbon interstitials ($I_C$). In this analysis, we focus specifically on the interactions of interstitials and Frenkel pair defects. As shown in Figs. 12 a-d, during the intermediate stage, different types of interstitial configurations, such as single, doublets, and triplets, are formed, along with Frenkel pair defects, which tend to migrate toward the void and accumulate either inside or at the void–matrix interface. This accumulation contributes to the shrinkage of the pre-existing void.

By the final stage, the number of interstitials and Frenkel pairs has significantly decreased, indicating that the void acts as a sink for point defect recombination. Therefore, while pre-existing voids facilitate the elimination of interstitials, they also alter the defect dynamics, favoring the formation of dispersed, smaller vacancy clusters.

The recombination efficiency ($\eta$), during the final stage, as a function of pre-existing void radius ($R_v$=5, 10, 15, and 20 Å), was calculated using Eq. (1) and is plotted in Fig. 12e. The plot shows that the value of ($\eta$) initially decreases and then saturates at around 36 ps. As highlighted in the inset of Fig. 12e, larger pre-existing voids correspond to higher recombination rates.

$$\eta = \left( \frac{N_{max} - N_d}{N_{max}} \right) \times 100 \tag{1}$$

where $N_d$ is the number of Frenkel pairs obtained at various times, and $N_{max}$ is the maximum number of Frenkel pairs obtained at t=0, during the final simulations.

In contrast, in the absence of voids, cascades tend to produce larger, more concentrated vacancy clusters, precursors to larger He-induced defects. Introducing voids via carbon implantation alters this trajectory by facilitating interstitial absorption and inhibiting cluster growth. The inward flux of interstitials into voids causes void shrinkage, while the corresponding vacancies remain dispersed within the lattice, often as isolated point defects or small clusters. This redistribution promotes the nucleation of helium bubbles rather than large platelets, consistent with TEM and DBS findings.

To gain deeper insight into the interaction between helium atoms and vacancy clusters in SiC, we calculated the monomer binding energy of He atom to a vacancy cluster ($V_n$), Helium ($He_m$), and mixed ($He_mV_n$) as shown in Fig. 13. Note that in all DFT calculations, the position of He interstitial is located at the R site, which represents the most stable interstitial position in 6H-SiC [61]. The monomer binding energy of a helium atom $E_b(He)$ to a Helium-vacancy complex containing $m$ He atoms and $n$ vacancies $He_mV_n$ was calculated using the following expression:

$$E_b(He) = E(He) + E(He_{m-1}V_n) - E(He_mV_n) - E(perfect) \tag{2}$$

Where $E(He)$ is the system's total energy with a single He atom occupying R interstitial site while $E(He_mV_n)$ and $E(perfect)$ representing the supercell's total energy with and without defects, respectively. Negative binding energy denotes an attractive interaction, and positive binding energy denotes a repulsive interaction.

Fig. 13a presents the DFT results of $E_b(\text{He})$ to vacancy clusters containing up to 16 mixed silicon $V_{Si}$ and carbon $V_C$ vacancies. In general, the data indicate that a He atom increasingly favors binding to a vacancy cluster as the cluster size grows, consistent with earlier DFT studies [57,61]. Moreover, the curve shows a pronounced initial increase and then reaches a plateau. The relative percentage difference (RDP) of $E_b(\text{He})$ is greatest during the early stages of clustering (the nucleation stage), and it decreases as the cluster size grows, resulting in lower RPD values for larger clusters. This trend indicates that He atoms strongly prefer to bind to small voids. In contrast, the binding of a He atom to pure He clusters is consistently repulsive, as shown in Fig. 13b, demonstrating that the formation of helium clusters in the absence of vacancies is energetically unfavorable. For mixed $\text{He}_m V_n$ clusters, two distinct trends emerge. When the He/V ratio is less than one, He atoms display a strong preference for binding to smaller clusters. However, as the He/V ratio approaches one, the attraction becomes noticeably weaker, as illustrated in Fig. 13c and 13d. These findings clearly show that helium atoms preferentially bind to small clusters, consistent with experimental observations of dispersed sub-nanometer He bubbles in pre-damaged samples.

The distinct behavior of He atoms with respect to He/V ratio highlights the critical role of cluster size: smaller clusters act as more effective trapping sites, promoting the formation of stable nanobubbles. In contrast, larger clusters, combined with higher helium concentrations and elevated temperatures that enhance local diffusion, favor helium accumulation and platelet formation.

Furthermore, we calculated the migration energy barrier for different $\text{He}_m\text{V}_n$ configurations to elucidate the tendency of He to form a bubble/platelet. Fig. 14 depicts the migration energy barrier results; it is found that He has the lower migration energy to bind to pure vacancy compared to He clusters, aligning with the corresponding monomer binding energy results, Fig. 13 a and b, indicating the importance of He-V complexes in bubble formation. Additionally, the He migration energy to $\text{He}_3\text{V}_3$ is lower than that of $\text{He}_3\text{V}_6$. When the He/V ratio is much greater than 1, the value of He migration energy is larger than that of $He_3V_3$ due to the increase in cluster size. This behavior is plausible at this ratio, as the formation of platelets is expected.

The correlation between simulation results and experimental observations highlights the crucial role of pre-existing damage in governing He-induced defect evolution in α-SiC. The behavior of helium in the matrix, with or without pre-damage, can be described as follows: During He implantation into pre-irradiated samples, collision cascades from incoming ions overlap with pre-existing voids. These voids serve as efficient sinks for interstitials, leading to localized void shrinkage (Fig. 12) and altering the cascade recombination dynamics. As a result, numerous small vacancy clusters form near the cascade core. These clusters serve as distributed trapping sites for helium, promoting a more spatially diffuse defect distribution. For instance, at 0.36 dpa, He bubbles are spread across a ~400 nm region, in contrast to the ~100 nm localization observed in undamaged samples, where helium accumulation leads to platelet and nanocrack formation. This spatial broadening is reflected in the wider strain profiles shown in Fig. 9. Pre-existing vacancy clusters thus play dual roles: (1) acting as sinks for mobile interstitials, and (2) serving as nucleation sites for helium trapping. The local He concentration in these small clusters remains below the threshold for platelet formation. Simulations show enhanced helium binding energies in smaller vacancy clusters, suggesting thermodynamic stabilization of dispersed nanobubbles over platelet formation. In contrast, in undamaged samples, cascade-induced clustering tends to produce larger vacancy aggregates. At high irradiation temperatures, enhanced mobility allows helium atoms to overfill these large clusters (He/vacancy ratio >1), triggering platelet nucleation and coalescence into over-pressurized platelets and eventually nanocracks (Fig. 2). Therefore, maintaining a high density of small, isolated vacancy clusters is key to suppressing extended defect growth. These findings underscore the potential of defect engineering to control the helium defect landscape. Low-dose pre-damage favors distributed He bubble formation, while higher-dose pre-damage leads to uniform arrays of sub-nanometer He bubbles with no evidence of platelets or cracks. This approach fundamentally alters helium behavior and enhances the radiation tolerance of α-SiC.

Although demonstrated here in α-SiC, the concept of defect landscape engineering is broadly applicable to a wide class of radiation-sensitive ceramics. Many high-performance ceramics used in nuclear, fusion, and aerospace environments, including carbides (ZrC, TiC, $B_4C$), nitrides (TiN, AlN, $Si_3N_4$), and oxides ($Al_2O_3$, MgO, $ZrO_2$), exhibit low He solubility and high mobility, leading to similar pathways of bubble coarsening, platelet formation, and crack initiation. The introduction of tailored pre-existing vacancy clusters offers a composition-independent lever for redirecting He transport and clustering, independent of grain size, phase composition, or interface density. In systems where compositional modification is impractical due to manufacturing constraints or qualification requirements, pre-damage engineering can be implemented through controlled ion implantation, neutron irradiation, or even additive manufacturing routes that introduce vacancy-rich regions during processing. The mechanistic principles identified here, vacancy clusters as dual-function helium traps and cascade sinks, are general and scalable, offering a versatile design pathway for extending the lifetime and reliability of ceramic components across multiple extreme-environment applications.

This work demonstrates that deliberately engineered vacancy landscapes can fundamentally reshape He defect evolution in ceramics. By stabilizing uniformly dispersed nanobubbles and suppressing extended defect formation, defect landscape engineering provides a general, composition-independent strategy for improving radiation tolerance. The approach reframes defect control in ceramics from a passive property to an actively tunable design parameter, enabling the creation of next-generation materials capable of sustained performance in the harshest operational environments.

**Materials and methods**

**Ion Distribution and Implantation Strategy**

To investigate the interaction of helium with pre-existing damage, sintered α-SiC samples were subjected to a series of controlled ion irradiations. Carbon self-ion irradiation was performed at room temperature using 1080 keV C ions at fluences of $2 \times 10^{15}$ ions/cm² and $4 \times 10^{15}$ ions/cm², resulting in displacement damage levels of approximately 0.25 dpa and 0.5 dpa, respectively. The use of carbon as a self-ion ensured no contamination by foreign elements and allowed precise control over defect generation; moreover, the energy is relevant to C recoil energies produced in SiC by fusion neutrons. Following C irradiation, the same samples were implanted with helium ions at 300 keV to a fluence of $1 \times 10^{17}$ He/cm². For comparison, separate samples were implanted solely with helium under identical conditions. During all irradiations, the samples were tilted 7° off normal incidence to minimize ion channeling effects. The resulting ion distributions and damage profiles were calculated using the SRIM code with the full-cascade simulations [62]. Displacement threshold energies of 20 eV and 35 eV were used for carbon and silicon atoms [63], respectively, assuming an atomic density of 3.21 g/cm³ for α-SiC. SRIM simulations indicate that the projected range (Rp) of helium implantation peaks at approximately 930 nm, with a maximum concentration of around 7 at%. Carbon implantation, on the other hand, results in peak damage levels centered at approximately 1060 nm. The corresponding C ion concentrations are relatively low, about 0.13 at% for the $2 \times 10^{15}$ ions/cm² fluence and 0.26 at% for the $4 \times 10^{15}$ ions/cm² fluence, with their maxima located deeper, around 1120 nm. To promote helium trapping in vacancy-rich regions while minimizing direct interaction with implanted carbon, the C implantation profile was designed to peak deeper than the He profile, so that migrating helium atoms primarily encounter pre-existing vacancy clusters rather than the dilute implanted C atoms. This intentional offset ensures that helium atoms primarily interact with damaged structures rather than with implanted carbon atoms themselves. As shown in Fig. 15, this design enables spatial separation between helium concentration and carbon content, thereby isolating the role of pre-existing damage in modifying helium behavior. This controlled implantation strategy enables a systematic study of how irradiation-induced defects influence helium trapping, diffusion, and cluster evolution in SiC.

**Helium Depth Profiling via ERDA**

To characterize the in-depth distribution of helium, Elastic Recoil Detection Analysis (ERDA) was conducted using a 48 MeV $Cl^{7+}$ ion beam. The incident beam was aligned at a 75° angle relative to the sample normal, with a scattering angle of 30°, covering an analysis area of approximately $2 \times 2$ mm². Recoiled atoms and scattered ions were detected using a Bragg Ionization Chamber (BIC), which enables both energy resolution and atomic number (Z) identification. Helium recoils were detected by a dedicated solid-state detector positioned at a 40° scattering angle. A 25 μm Kapton foil was placed in front of this detector to block heavier recoils and scattered ions, ensuring selective helium detection. Although this setup reduces depth resolution due to energy loss straggling in the foil, it compensates by allowing deeper probing, critical for profiling He implanted at an average depth of ~1 μm. To enhance analysis depth, ion beam energies of 48 MeV and 47 MeV were employed. However, at such high energies, recoiled carbon atoms may also penetrate the foil, contributing to background signals in the helium spectra. Beam dose was monitored using a gold-plated rotating vane (1 Hz) and a solid-state detector that recorded Cl backscattering from the gold target. All data, including BIC events, helium detector signals, and rotating vane counts, were recorded in list mode. This format allowed time-resolved monitoring of elemental loss during ion bombardment, particularly helium depletion. Post-acquisition, the helium loss behavior was evaluated by plotting the total helium counts against the accumulated dose. These plots were then fitted and extrapolated to zero dose to estimate the original elemental concentrations prior to any ion-induced depletion. The collected data were analyzed using NDF software version 9.6i [64].

**TEM and Strain Mapping via Nano-Precession Electron Diffraction (N-PED)**

TEM measurements were carried out to investigate the microstructure of the samples. Cross-sectional TEM lamellae were machined by focused ion beam (FIB) using an FEI Helios NanoLab 660 workstation. , with a Ga ion source capable of energies up to 30 kV. Trenches were cut at 9 nA, and cross sections were cleaned at 2 nA. Then, the lamella was cut and transferred to a Cu TEM grid. The lamella was polished using accelerating voltages from 30 to 2 kV and ion currents ranging from 20 nA to 7 pA. After preparation, the TEM samples were imaged using both a JEM-2200 FS TEM and FEI Tecnai TF20 X-Twin operated at 200 kV using high-angle annular dark-field (HAADF) and bright-field (BF) detectors. TEM BF images were acquired at under- and over-focused conditions with a defocus value of 400 nm to observe the platelets, nanocracks, and bubbles present in the samples. STEM-EELS spectra were acquired using a Gatan GIF 2001 system with an energy resolution of ~0.7 eV, a convergence semi-angle of 5.86 mrad, a collection semi-angle of 14.68 mrad, a beam current of 9.87 nA, and a probe size of ~0.35 nm. The acquisition conditions were optimized to enhance the C K-edge (onset ~284 eV) and Si $L_{2,3}$-edge (onset ~99–100 eV) signals while minimizing sample drift.To resolve strain fields induced by ion implantation, Nano-Precession Electron Diffraction (N-PED) was performed using a Tecnai TF20 TEM operating at 200 kV, equipped with a NanoMEGAS DigiSTAR precession system and controlled by Topspin software [65]. A 20 μm condenser aperture was used to produce a quasi-parallel electron beam, resulting in high-contrast diffraction spots. The electron beam was processed at a 1° angle and scanned across the irradiated region with a step size of 5 nm and a beam diameter of 4 nm. Diffraction patterns were acquired at each position with ~20 ms exposure. Reference patterns were collected from a nominally strain-free region to serve as the baseline ($\varepsilon_{zz} = 0$). Cross-correlation algorithms were applied to extract spot shifts and compute the 2D strain tensor. Using the ASTAR platform and AutoSTRAIN module, strain and orientation maps were generated with nanometer spatial resolution. The beam precession technique minimizes dynamical scattering effects and reduces contrast artifacts due to sample thickness variations, significantly improving the accuracy and fidelity of strain measurements, and this method is used to determine local strain around grain boundaries, dislocations, and twins[66–70]. Virtual bright-field STEM images were also produced by digitally selecting the central diffraction spot, providing a complementary structural contrast map of the irradiated zone.

**Depth-resolved Doppler broadening measurements**

DBS was performed using a continuous monoenergetic beam, capable of positron implantations in the range of E = 0.04-35 keV. A high-purity Ge detector, with an energy resolution of 1.09 ± 0.01 keV at 511 keV, was used to record the characteristic 511 keV gammas at each beam energy. The experimental data were interpreted by measuring the energy deviation in the characteristic 511 keV annihilation gammas induced by the Doppler shift of the low-momentum valence electrons and were represented in terms of the line-shape defect parameter known as the S parameter, a fraction of the spectrum in the middle region (511±0.70 keV). Details about DBS spectroscopic measurements and analysis can be found elsewhere [71,72].

**Simulation methods**

We carried out a series of cascade simulations to understand the impact of pre-existing damage on the radiation-induced defects. The 6H-SiC crystal is a hexagonal crystal structure with a lattice constant of a = b = 0.3081 nm and c = 1.5117 nm. The simulation cell size was constructed by repeating the 6H unit cell by 60 × 90 × 20 along x, y, and Z directions, containing about 1,100,000 atoms. The simulation box is oriented in the x[1100], y[1120], and z[0001] directions. The obtained cells were initially quenched to 0K to eliminate the excess potential energy and then relaxed at 1073 K using an isobaric-isothermal Berendsen thermostat in 200 ps. During cascade simulations, we adopted a multi-phase timestep procedure as follows: A variable time step with a maximum value of 1fs was used following the procedure described in Refs [19,73]. The system size is sufficiently large so that the system temperature did not appear to rise during the simulations, indicating that the cascade defect formation was slightly influenced and the thermal bath was not considered. The cascade was established by assigning a kinetic energy of 10 keV to a selected Si atom as a primary knock-on atom (PKA) along the negative z-direction. The void with a prescribed radius was generated by removing the atoms from a sphere whose center is located at the center of the box. All MD simulations were carried out using a large-scale atomic/molecular massively parallel simulator (LAMMPS) developed by Plimpton et al. [74]. The interaction between Si-C atoms was modeled by the Tersoff-type potential developed by Devanathan et al.[75]. The short-distance interaction was modeled using the ZBL (Ziegler-Biersack-Littmark) [76], which accounts for the short-range interaction occurring during cascade simulations. The Wigner–Seitz analysis was used to quantify the interstitials and vacancies in the system, and cluster analysis was used to determine the size of newly generated vacancy clusters as implemented in OVITO [77].

We investigated the energetics (binding and migration) of various defects using the Vienna Ab initio Simulation Package, VASP [78]. Exchange and correlation functions were taken in a form proposed by Perdew and Wang (PW91) within the generalized gradient approximation (GGA) [79]. The supercell size is 4×4×1 unit cell containing 192 atoms. The energy cutoff for the plane-wave basis set used throughout this work is 450 eV for relaxation of the atomic position, shape, and volume of the supercell. For Brillouin zone integration, a uniform k-point mesh of 5×5×5 was employed using the Monkhorst-Pack scheme. The migration barrier energy of a defect was calculated using the climb image nudged elastic band CI-NEB [80] method implemented in the VTST code [81]. For NEB simulation, a smaller 3×3×1 supercell was used, and a minimum of nine images along the migration path was adopted. During minimization, all atoms within the supercells are allowed to undergo full relaxation. Atomic positions within the supercells are relaxed with tolerances of 0.01 eV/Å for atomic force and $10^{-6}$ eV for total energy.

**Data availability**

The data that support the findings of this study are available within the article and its Supplementary Information files. Additional datasets generated and/or analyzed during the current study are available from the corresponding authors upon reasonable request.

**Code availability**

The Vienna *Ab initio* Simulation Package (VASP) is available at https://www.vasp.at/. LAMMPS Molecular Dynamics Simulator (LAMMPS) is available at https://www.lammps.org/. Open Visualization Tool (OVITO) is available at https://www.ovito.org/. The custom codes used in this work are available from the corresponding authors upon reasonable request.

**Acknowledgements**

This work was financially supported by the European Union under the project Robotics and advanced industrial production (Reg. No. CZ.02.01.01/00/22_008/0004590). The authors also disclose support for the research of this work from the Scientific Grant Agency of the Ministry of Education, Science, Research and Sport of the Slovak Republic and the Slovak Academy of Sciences (VEGA grant No. 1/0511/25). A. M acknowledges the assistance provided by the Advanced Multiscale Materials for Key Enabling Technologies project, supported by the Ministry of Education, Youth, and Sports of the Czech Republic. Project No. CZ.02.01.01/00/22_008/0004558, Co-funded by the European Union." – AMULET project. Parts of this research were carried out at IBC at the Helmholtz-Zentrum Dresden-Rossendorf e. V., a member of the Helmholtz Association. Sichuan Science Technology Program (Grant No. 2025NSFJQ0044). The National Natural Science Foundation of China (No. 12575304). Parts of this research were carried out at ELBE at the Helmholtz-Zentrum Dresden–Rossendorf. V., a member of the Helmholtz Association. We would like to thank the facility staff for their assistance. This work was supported by the Ministry of Education, Youth and Sports of the Czech Republic through the e-INFRA CZ (ID:90254).

**Author contributions**

Nabil Daghbouj performed TEM lamella preparation, TEM imaging, strain mapping, data curation, formal analysis, visualization, and contributed to conceptualization, methodology, project administration, and writing of both the original draft and the revised manuscript. Ahmed T. AlMotasem carried out the DFT simulations, data curation, formal analysis, visualization, and contributed to methodology and manuscript writing. Bingsheng Li performed the ion irradiation experiments and contributed to manuscript revision. Vladimir Krsjak contributed to visualization, validation, and manuscript revision. Jan Duchon conducted the STEM–EELS measurements. Fang Ge contributed to manuscript review and editing. Maciej Oskar Liedke and Andreas Wagner performed the DBS measurements and contributed to manuscript revision. M. Bensalem contributed to manuscript review and formal analysis. Fateh Bahadur prepared TEM lamellae. Frans Munnik performed the ERDA measurements and contributed to manuscript revision. Miroslav Karlik and Anna Mackova contributed to visualization, validation, and manuscript revision. Tomas Polcar provided supervision, project administration, and funding acquisition, and contributed to visualization, validation, and manuscript revision. William J. Weber contributed to visualization, validation, formal analysis, and manuscript revision.

**Competing interests**
The authors declare no competing interests.

**List of captions**

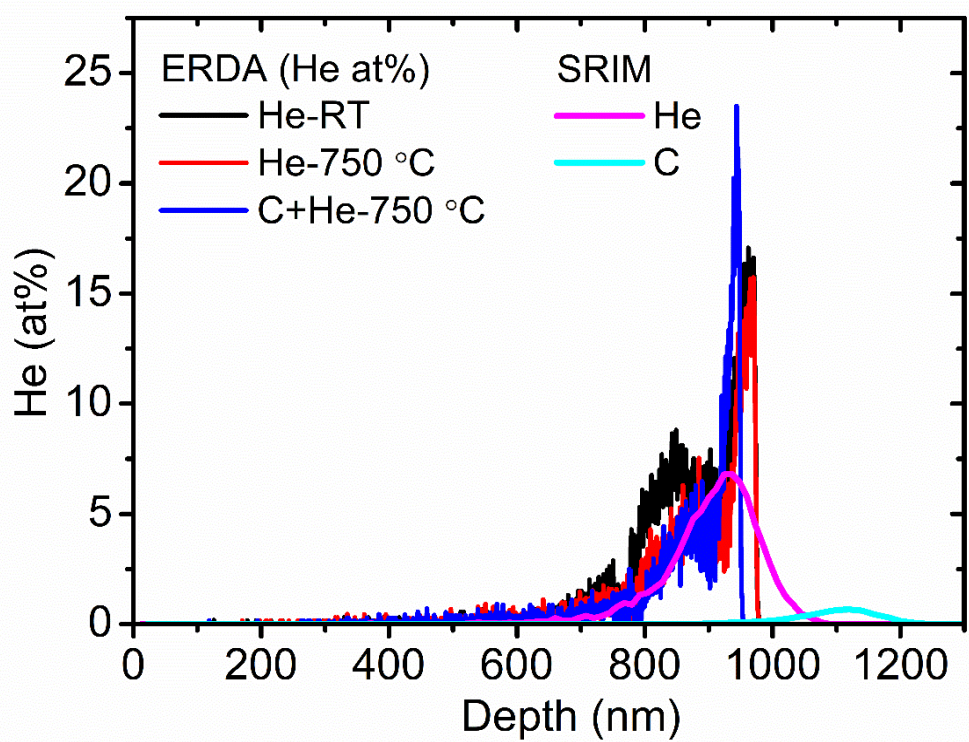

**Fig. 1. Tailored pre-damage alters helium depth distribution in α-SiC.** He depth profiles from ERDA for samples implanted at RT, 750 °C, and co-implanted in pre-damaged SiC (0.36 dpa) at 750 °C. Two Gaussian peaks obtained from ERDA indicate He redistribution.

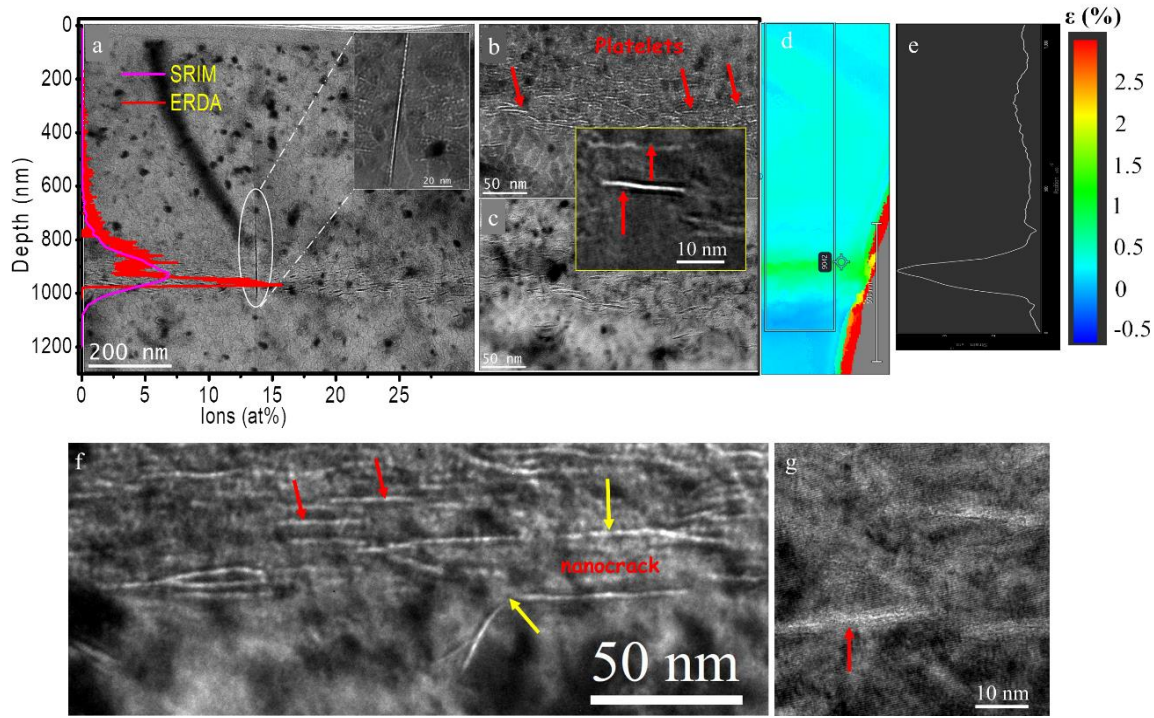

**Fig. 2. High-temperature helium implantation produces extended defects in undamaged α-SiC.** TEM characterization of sintered α-SiC implanted with $1 \times 10^{17}$ He/cm² at 750 °C: (a) Overview showing helium platelet formation overlaid with ERDA and SRIM helium depth profiles. The inset highlights a zone depleted of bubbles/platelets near the GBs; (b, c) Bright-field TEM images acquired under under-focus (b) and over-focus (c) conditions, showing He platelets marked by red arrows at higher magnification. The inset presents two He platelets in detail; (d) Strain map obtained by N-PED along the [01-1-4] zone axis, revealing ~1% tensile strain within the grain interior and ~2.5% strain near and along the GB; (E) Corresponding strain profile extracted from the rectangular region in (d); (f, g) High-magnification TEM images showing nanocracks (yellow arrows) and elongated helium platelets (highlighted with red arrows) aligned with the peak helium concentration; (g) Lattice-resolution micrograph showing platelets with helium gas trapped between distorted atomic planes.

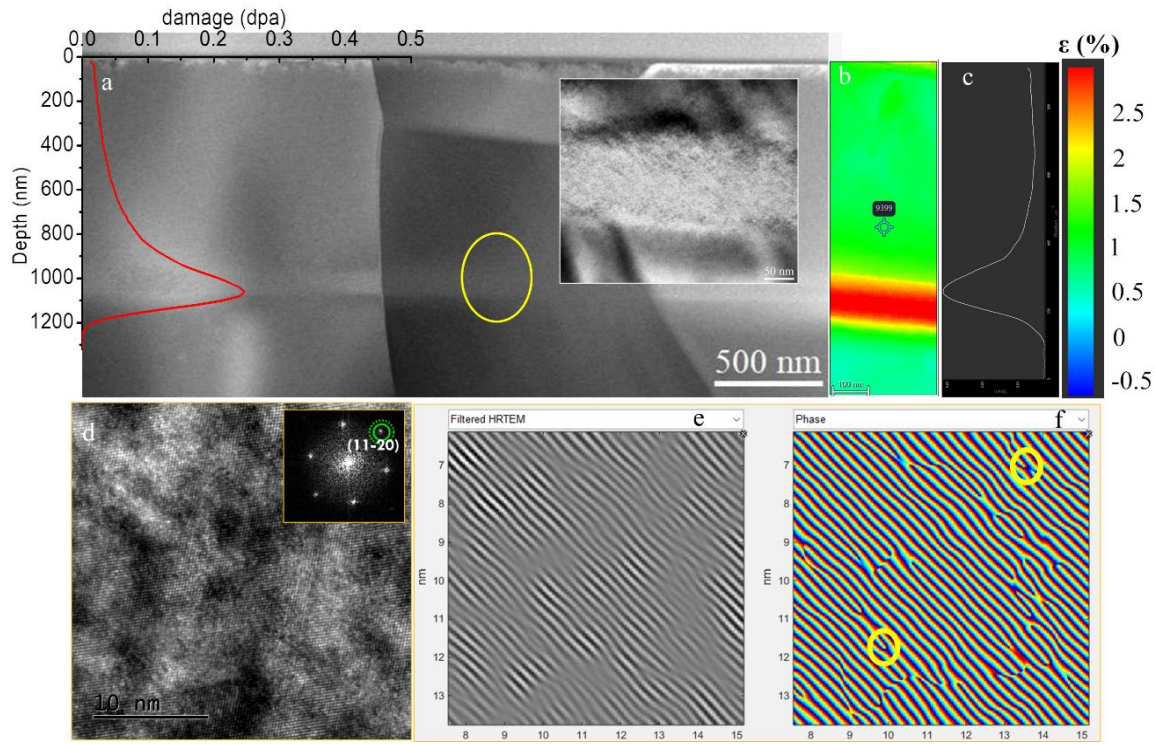

**Fig. 3. Microstructural and strain effects of carbon irradiation at 0.25 dpa in α-SiC.** Microstructural and strain analysis of sintered α-SiC pre-damaged to 0.25 dpa with carbon ions : (a) BF-STEM image showing the carbon damage profile peaking at ~1060 nm; inset highlights the defect-rich region; (b) Strain map acquired via N-PED along the [01-1-4] zone axis, revealing a maximum tensile strain of ~3% at the peak damage depth; (c) Strain profile corresponding to the region indicated in (b); **(d)** HRTEM image viewed along the [$\bar{2}$203] zone axis. The inset shows the corresponding fast Fourier transform (FFT); the (11$\bar{2}$0) reflection used for the geometric phase analysis is marked by a green circle. **(e)** Inverse-FFT filtered HRTEM image obtained by selecting the (11$\bar{2}$0) reflection. The image reveals well-resolved (11$\bar{2}$0) lattice fringes with local distortions. **(f)** Corresponding phase map for the (11$\bar{2}$0) g-vector. Phase is approximately linear within undistorted regions but shows pronounced local perturbations and "necklace-like" features highlighted by yellow circles, indicating lattice displacement gradients.

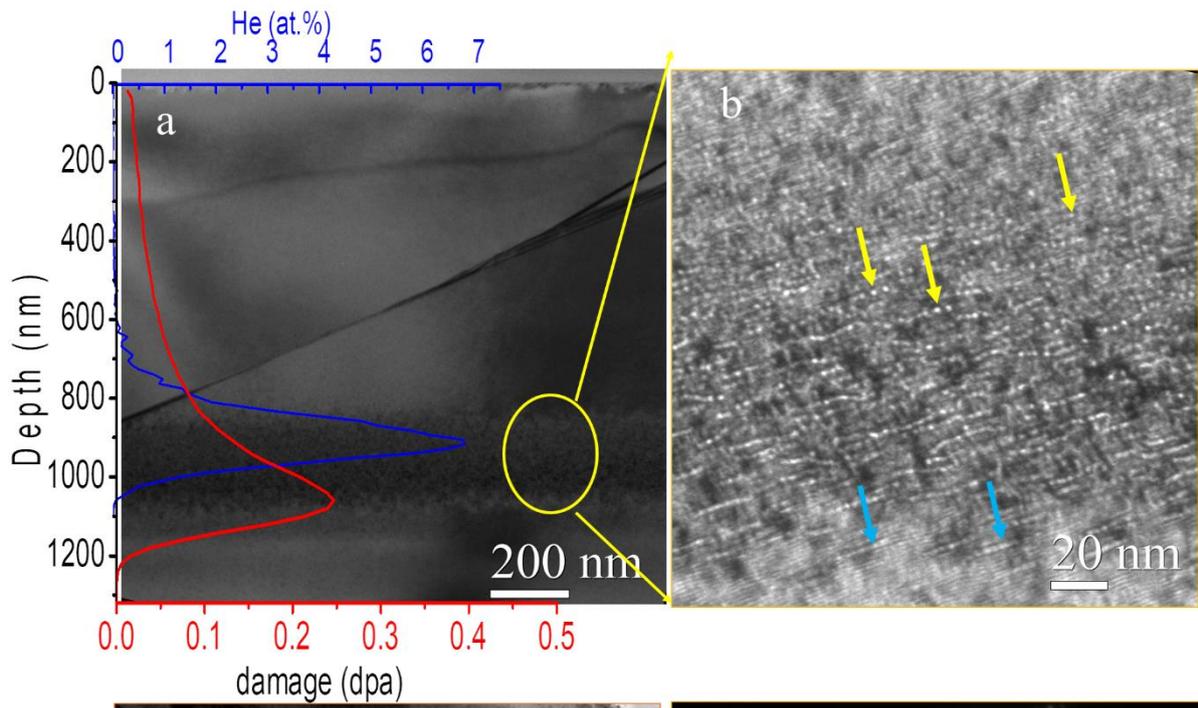
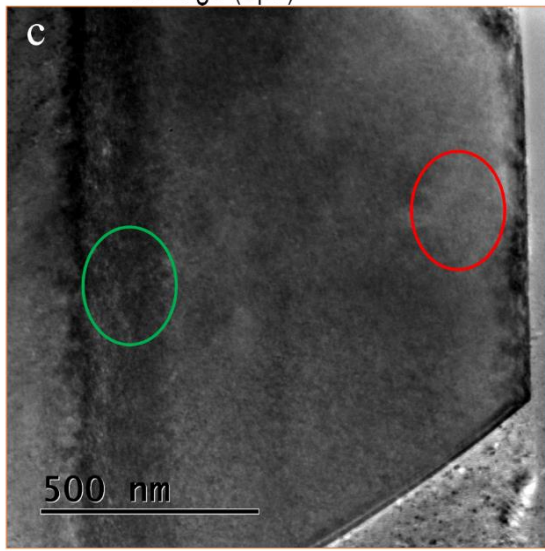
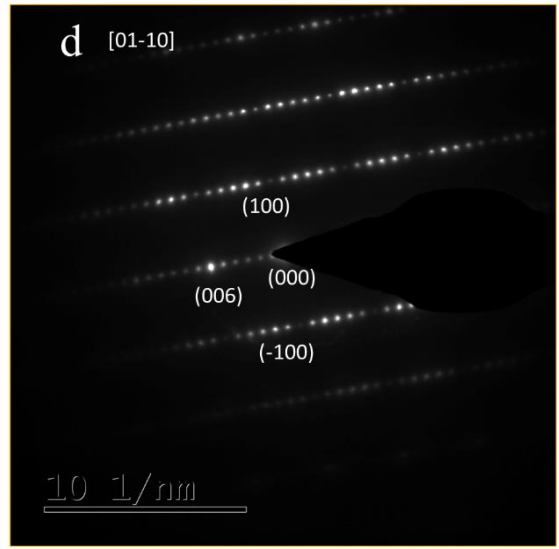
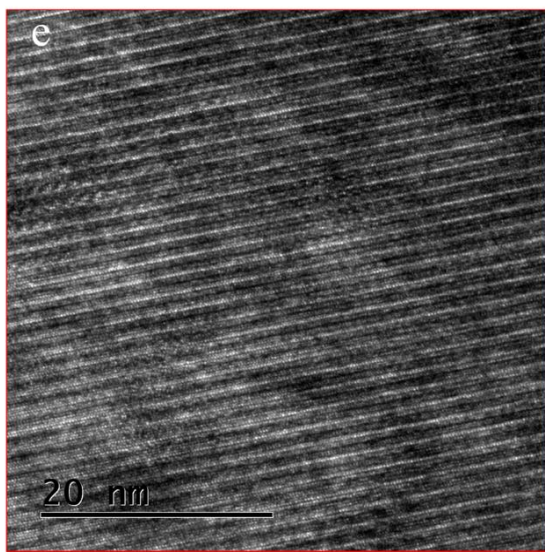
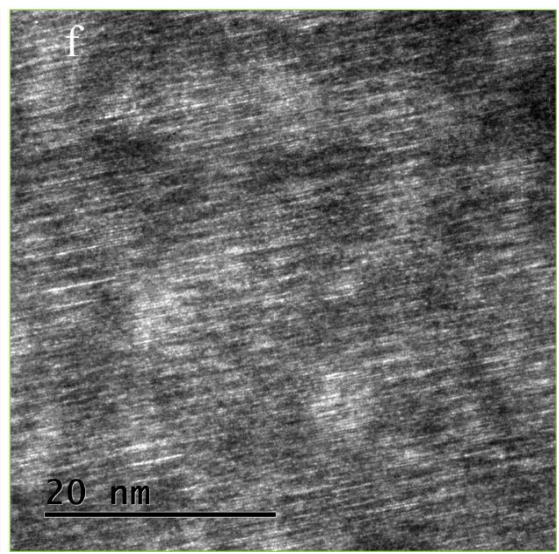

Fig. 4. **Low-dose pre-damage suppresses platelet formation and promotes bubble arrays.** (a) BF-TEM image of sintered α-SiC pre-damaged to 0.25 dpa with carbon ions and then implanted with helium at 750 °C ($1 \times 10^{17}$ He/cm$^2$), showing both helium and carbon damage profiles. (b) High-magnification view of the area marked with a yellow ellipse in (a), revealing discrete helium bubbles (yellow arrows) and bubble arrays (blue arrows), with complete suppression of platelet formation. (c) Low-magnification BF-TEM image of the grain oriented along the zone axis [01$\bar{1}$0]. (d) Selected-area electron diffraction pattern indexed for the [01−10]zone axis, confirming the single-crystal nature and orientation of the analyzed region. (e) HRTEM image taken near the surface, highlighted by a red circle. (f) HRTEM image taken at the peak damage region, highlighted by a green circle.

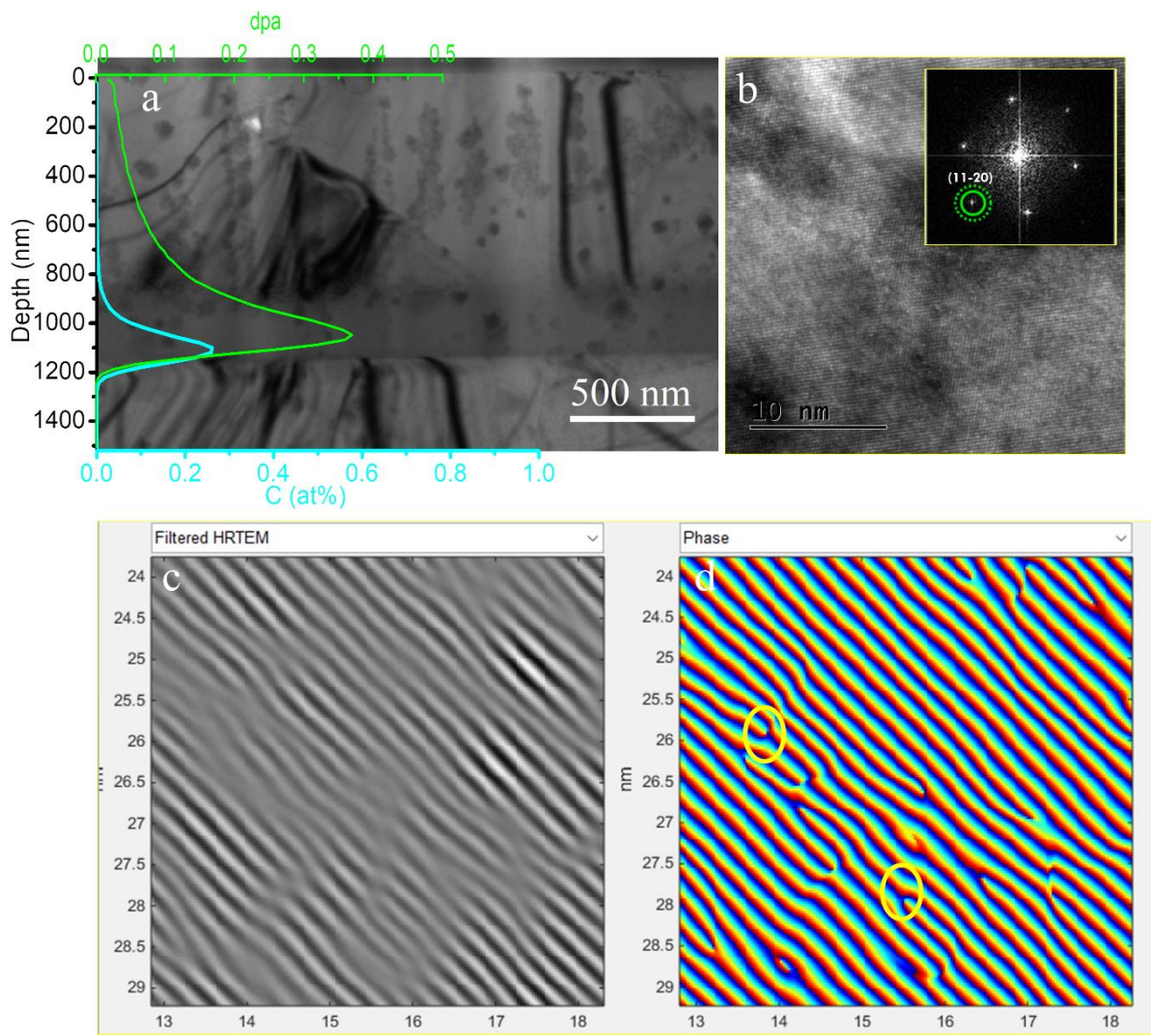

Fig. 5. **Microstructural and strain effects of carbon irradiation at 0.36 dpa in α-SiC.** Microstructural and strain analysis of sintered α-SiC pre-damaged to 0.36 dpa with carbon ions and subsequently implanted with helium at 750 °C: (a) BF-STEM image showing the carbon

damage profile peaking near 1060 nm; **(b)** HRTEM image viewed along the $[\bar{2}203]$ zone axis. The inset shows the corresponding fast Fourier transform (FFT); the $(11\bar{2}0)$ reflection used for the geometric phase analysis is marked by a green circle. **(c)** Inverse-FFT filtered HRTEM image obtained by selecting the $(11\bar{2}0)$ reflection. The image reveals well-resolved $(11\bar{2}0)$ lattice fringes with local distortions. **(d)** Corresponding phase map for the $(11\bar{2}0)$ g-vector. Phase is approximately linear within undistorted regions, but shows pronounced local perturbations and "necklace-like" features, highlighted by yellow circles, indicating lattice-displacement gradients.

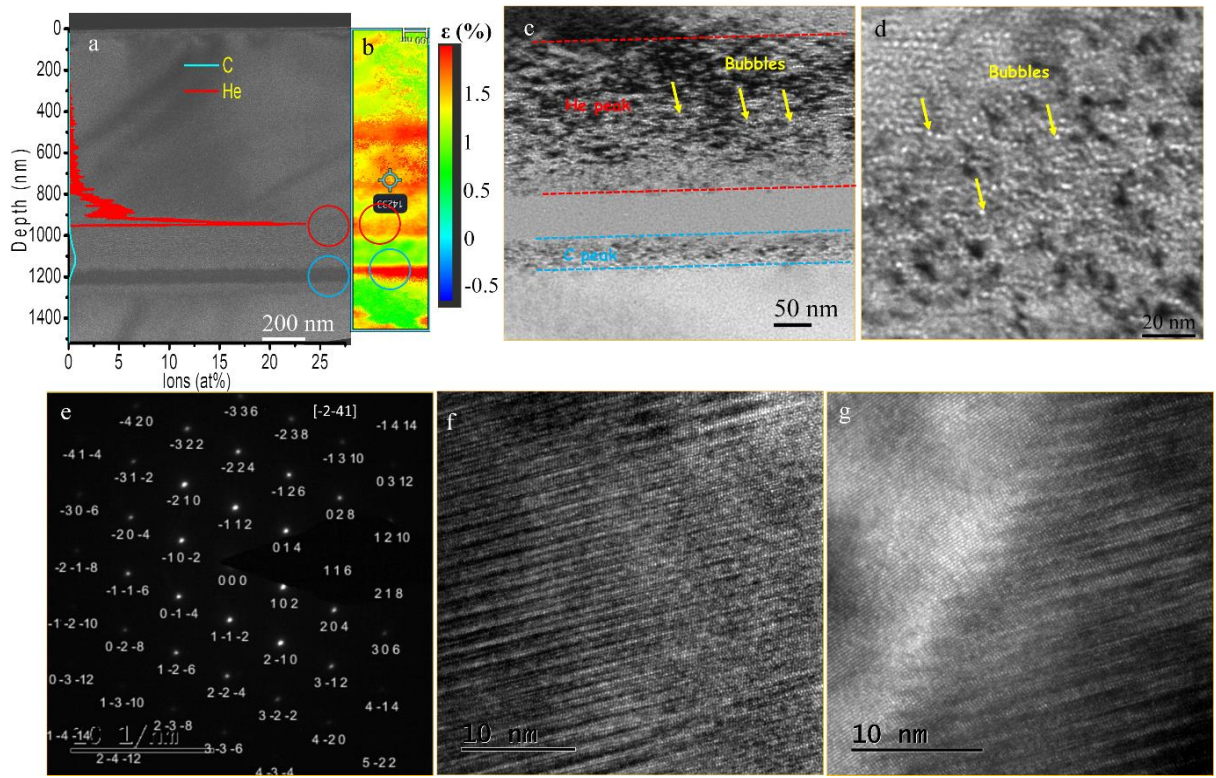

Fig.6. **High-dose pre-damage produces only nanoscale helium bubbles and reduces strain.** (a) BF-TEM image after co-implantation with $1 \times 10^{17}$ He/cm² at 750 °C, with both helium and carbon profiles overlaid; (c) N-PED strain map along the [214] zone axis, showing a maximum tensile strain of ~1.5% at the peak damage depth; (d, e) High-magnification TEM images of the region marked in (b), taken under under-focus (d) and over-focus (e) conditions, showing nanoscale helium bubbles (~1 nm, yellow arrows) with no evidence of platelet or nanocrack formation. Inset in (d) provides an enlarged view confirming the absence of extended defects. (e) SADP indexed for the [214] zone axis, confirming the orientation of the analyzed region. (f) HRTEM image taken near the surface, highlighted by a red circle. (g) HRTEM image taken at the peak damage region, highlighted by a green circle.

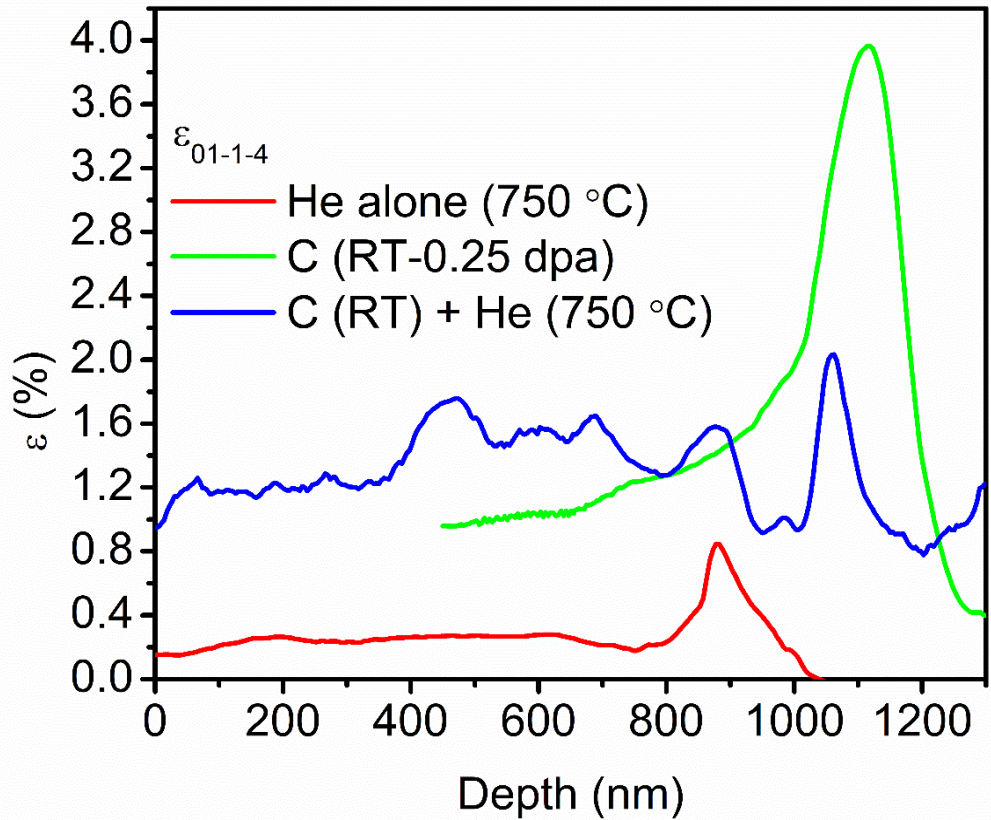

**Fig. 7. Pre-damage level controls strain magnitude and distribution.** Comparison of strain profiles $\varepsilon_{01\text{-}1\text{-}4}$ measured perpendicular to the grain surface under three irradiation conditions: Helium implantation at 750 °C without pre-damage (red curve); Carbon pre-damage to 0.25 dpa followed by helium implantation at 750 °C (green curve); Carbon pre-damage to 0.36 dpa followed by helium implantation at 750 °C (blue curve).

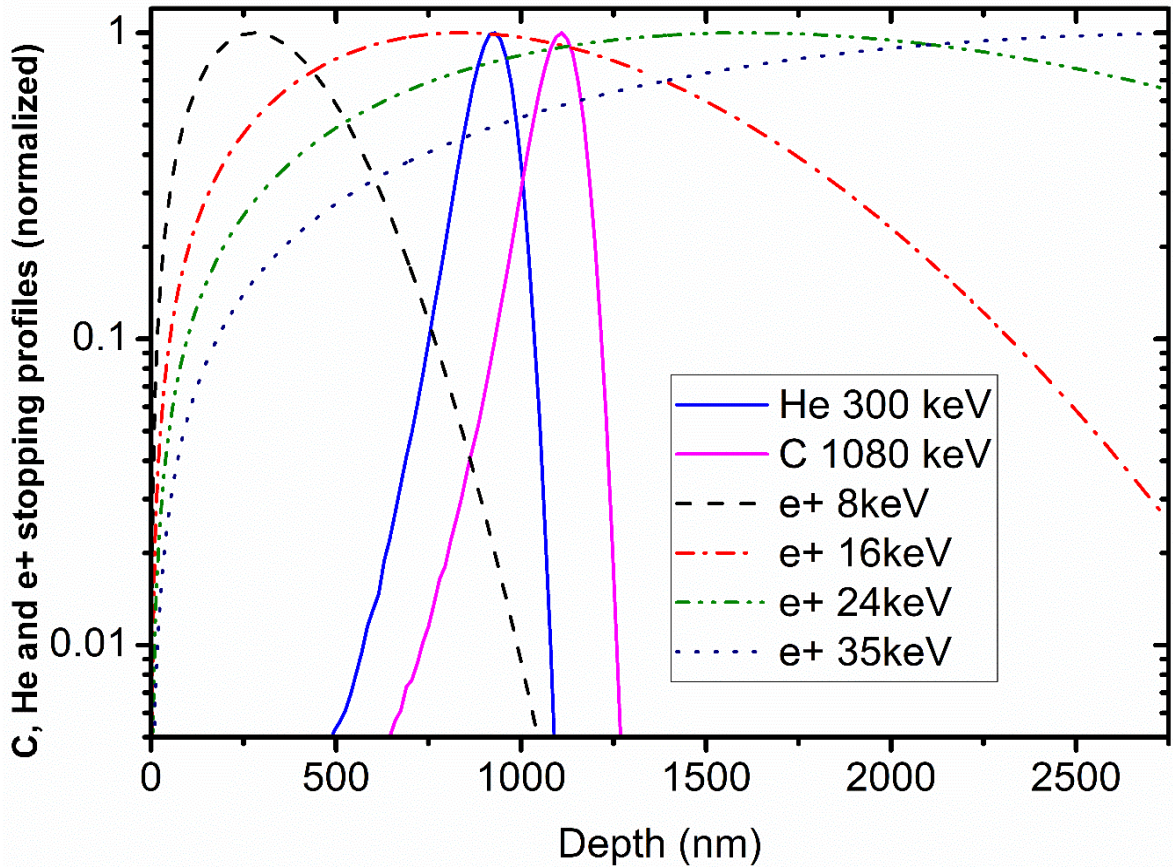

**Fig. 8. Depth sensitivity of positron beam analysis in irradiated α-SiC.** Stopping profiles of monoenergetic positrons and implanted ions in α-SiC, illustrating depth-dependent sensitivity of slow positron beam analysis. The figure includes stopping profiles for 300 keV He ions, 1080 keV C ions, and positrons with varying implantation energies.

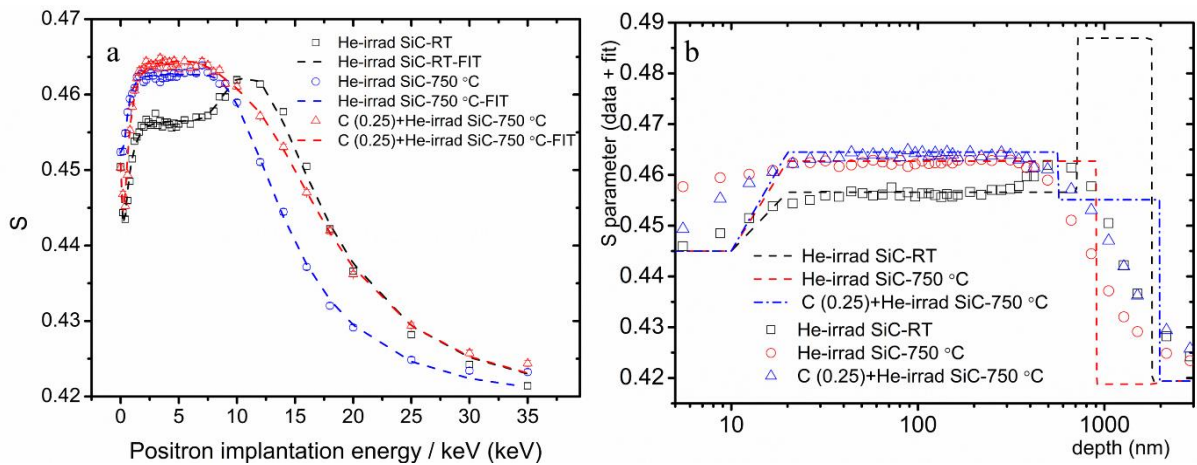

**Fig. 9. Pre-damage alters vacancy defect populations detected by positron annihilation spectroscopy.** S parameter depth profiles from Doppler Broadening Spectroscopy (DBS) and corresponding VEPFIT models for SiC samples implanted at RT, and 750 °C. (a) S vs. positron

energy and (b) reconstructed S vs. depth reveal distinct defect distributions for He and C+He co-implantation, with deeper and more pronounced damage in the dual-ion case

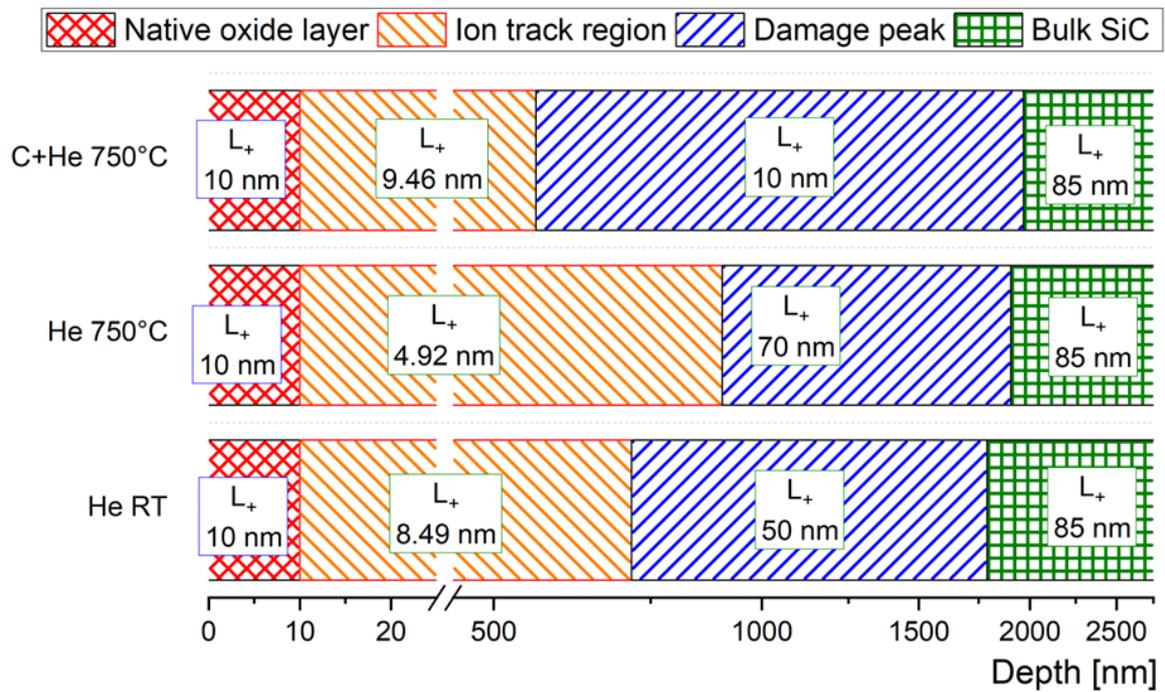

**Fig. 10**. **Layer model from VEPFIT analysis reveals depth-dependent defect structure.** Schematic representation of the fitted layer structure and positron diffusion lengths ($L^+$) obtained from VEPFIT analysis of slow positron Doppler broadening measurements on He-irradiated and C+He co-irradiated α-SiC samples

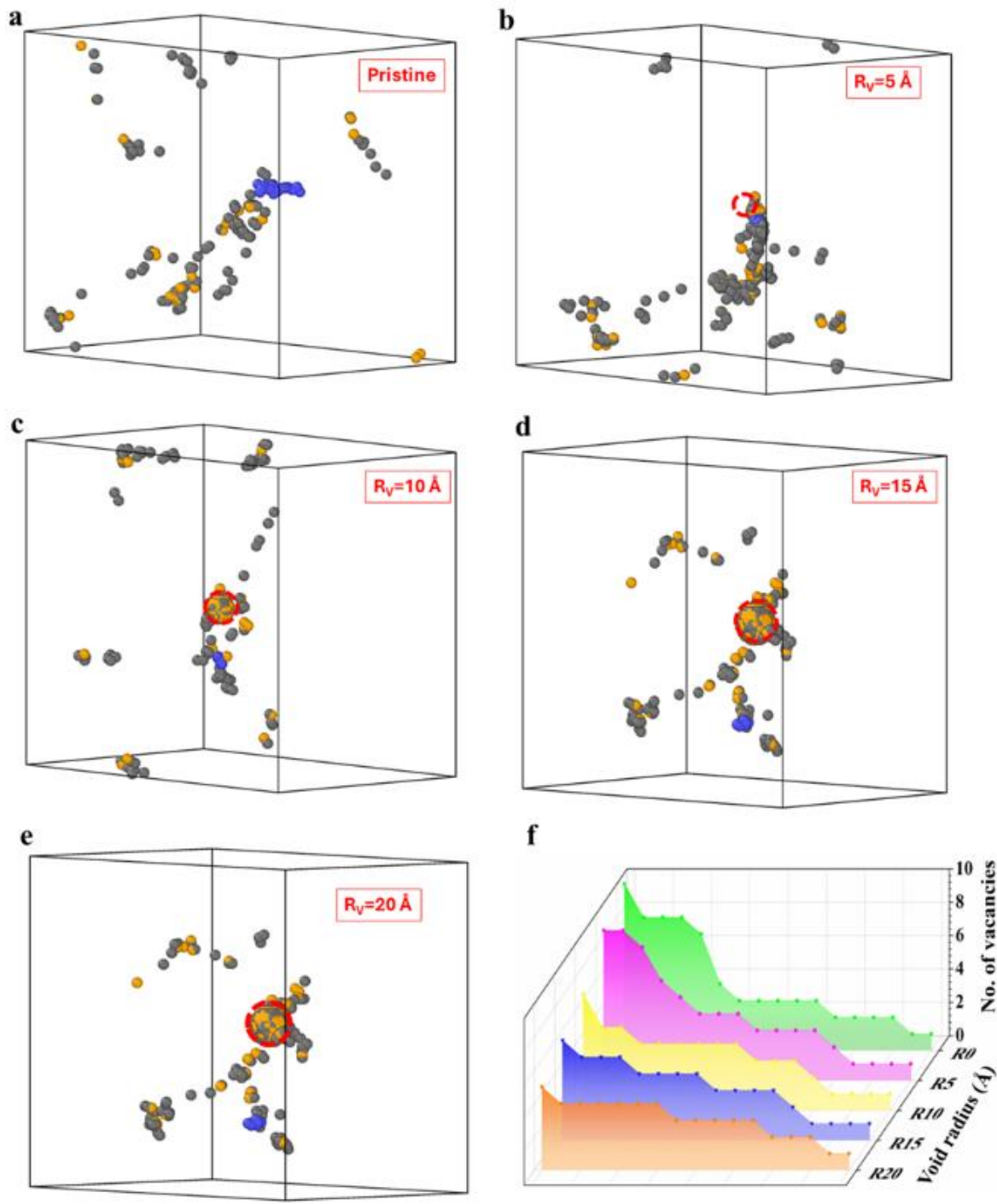

**Fig. 11. Molecular dynamics simulations show void-mediated suppression of large vacancy clusters.** MD simulation snapshot: cascade interactions with pre-existing voids in SiC. Larger vacancy clusters are suppressed in favor of distributed small clusters. Orange and grey spheres represent $V_{Si}$ and $V_C$, respectively. Red dashed circle denotes the pre-existing void, and blue color highlights the largest cluster size. The Wigner-Sietz analysis was used to quantify cascade defects.

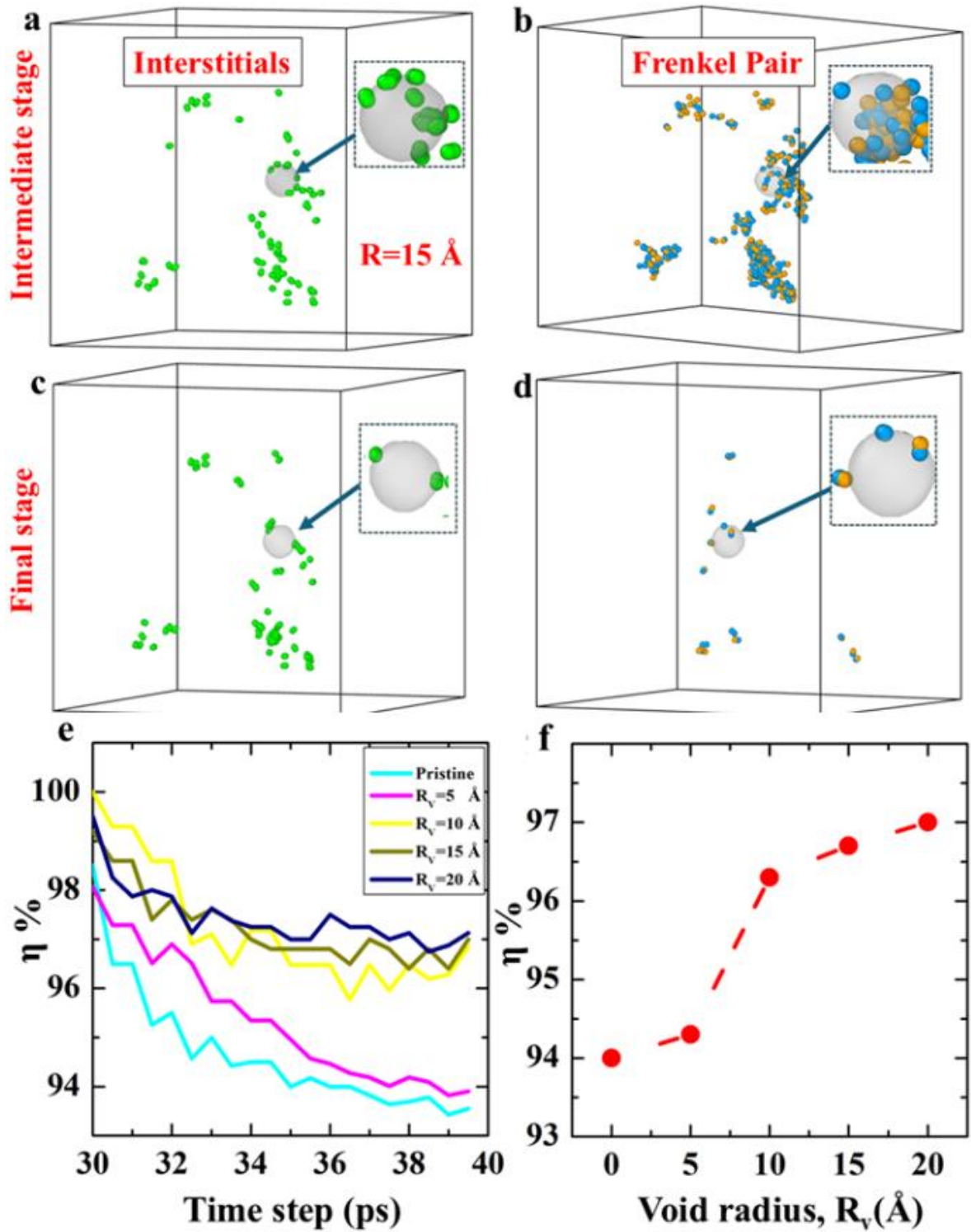

**Fig. 12. Pre-existing voids act as dynamic sinks for cascade-generated interstitials.** Atomistic snapshots showing the evolution of interstitial-type defects (a, c) and corresponding Frenkel pair defects (b, d) at intermediate and final stages for $R_v$ = 15 Å. (e) Plot of the recombination efficiency, Eq.(1), as a function of time during final stage of cascade

simulations. (f) the plot of the average value of recombination efficiency (over the last 4 ps) for different void radius. Green and orange(cyan) colors denote interstitial, vacancy(interstitial) Frenkel pairs.

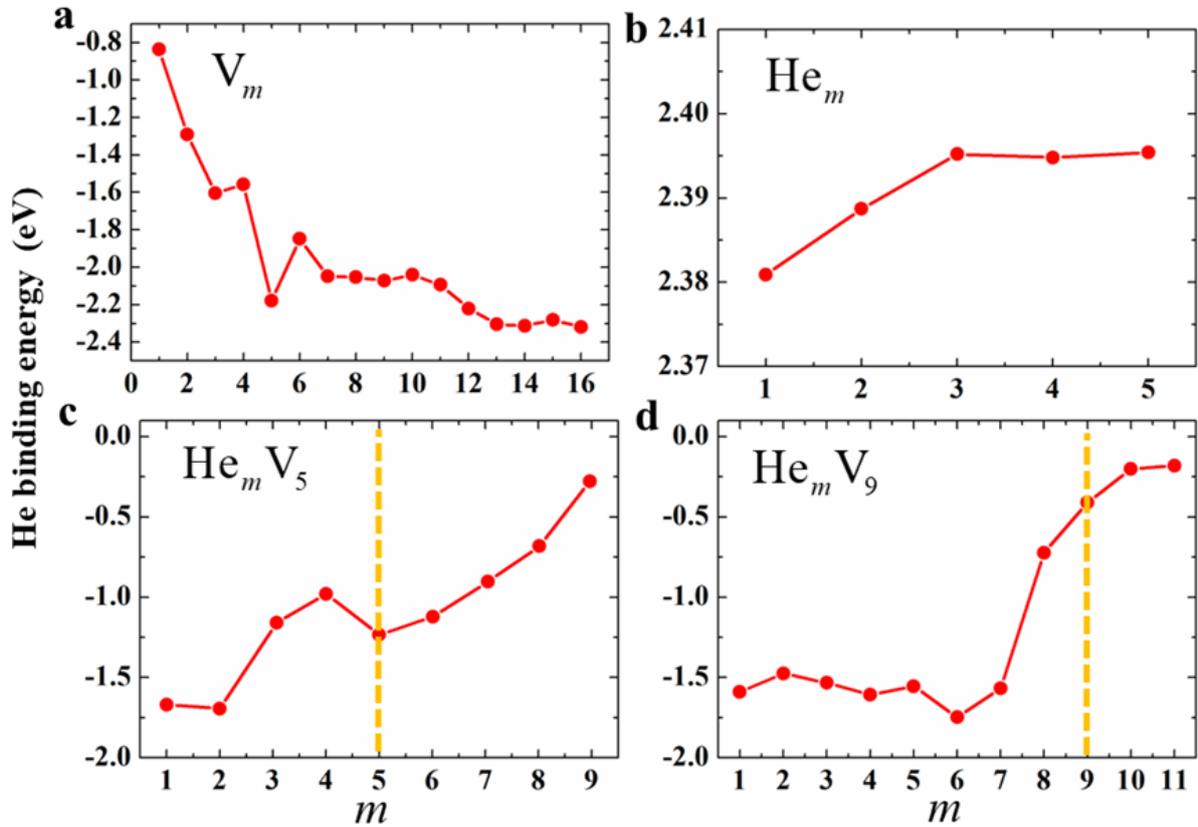

**Fig. 13. DFT binding energy calculations show strong helium affinity for small vacancy clusters.** DFT results of binding energy, calculated using Eq. (2), of the He atom to various pure $He_mV_n$ clusters (a, b) and mixed $He_mV_n$ clusters (c, d). Dashed yellow denotes He/V=1. Stronger binding for small vacancy clusters supports the experimental observation of platelet suppression.

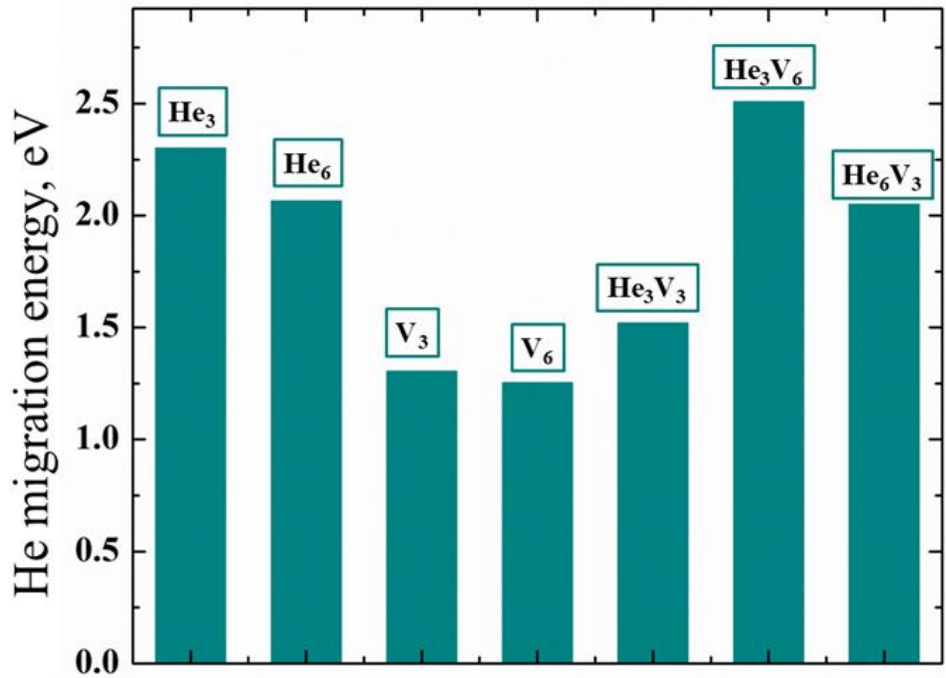

**Fig. 14. Helium migration barriers confirm preference for vacancy trapping over platelet growth.** Migration barrier energy (eV) for the He atom to various $He_mV_n$ clusters. Migration energy calculations show that He atoms migrate more readily to vacancies than to existing helium clusters. Lower migration barriers for small $He_mV_n$ complexes indicate a kinetic preference for nanobubble formation, whereas high He/V ratios favor platelet nucleation.

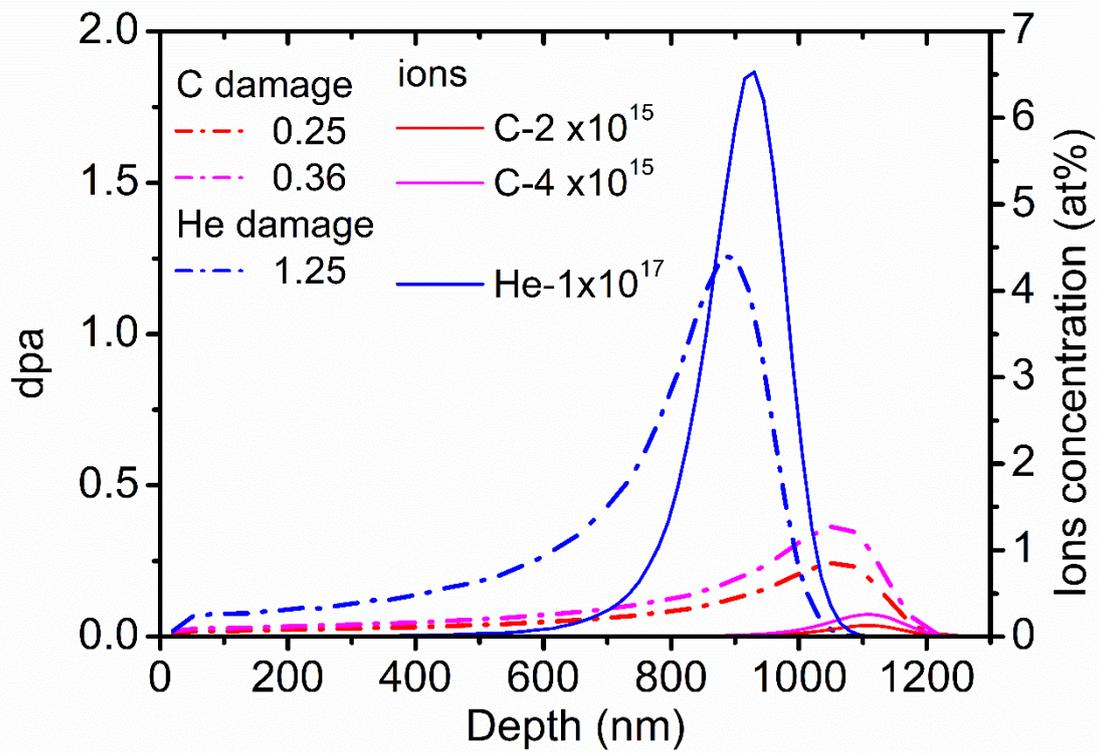

**Fig. 15.** SRIM-calculated profiles of implanted He and C ions in α-SiC. The C ion damage (0.25 and 0.36 dpa) is engineered to peak beyond the He implantation region to avoid direct overlap and to facilitate He interaction primarily with damage.